\documentclass[%
 reprint,
 amsmath,amssymb,
 aps,
 prd,
 longbibliography,
]{revtex4-2}
\usepackage{graphicx}% Include figure files
\usepackage{dcolumn}% Align table columns on decimal point
\usepackage{bm}% bold math
\usepackage{multirow}
\usepackage{color}

\usepackage[colorlinks,citecolor=blue]{hyperref}

\usepackage{float}
\usepackage{rotating}
\usepackage{upgreek}%for a charactor in a equation
\usepackage{subcaption}
\usepackage[normalem]{ulem} %for strikethought package 
\usepackage{comment}

\newcommand{\dmbaratm}{$\Delta \overline{m}^2_{31}$}

\newcommand{\dcptdm}{$\delta_{\nu\overline{\nu}}(\Delta m^2_{31})$}

\newcommand{\nova}{NO$\nu$A}

\begin{document}

%\preprint{APS/123-QED}
%\title{Testing $CPT$ Symmetry via Atmospheric Mass-squared Splitting from Long-Baseline Neutrino Oscillation Experiments}
\title{Atmospheric Mass-Squared Splitting at Sub-Percent Precision as a $CPT$ Symmetry Probe}

%\textsuperscript{*}
\author{T. V. Ngoc$^{1}$}
%\thanks{Corresponding Author tran.ngoc.64m@st.kyoto-u.ac.jp}
%\thanks{Corresponding Author tranngocapc06@ifirse.icise.vn}
\thanks{Corresponding Author tranngocapc06@gmail.com}
\author{S. Cao$^{1}$}%
%\email{cvson@post.kek.jp}
%\author{T. Nakaya$^{3}$}
%\author{N. T. Hong Van$^{3}$}
%\email{nhvan@iop.vast.ac.vn}
\author{P. T. Quyen$^{1,2}$}
%\email{phantoquyen@ifirse.icise.vn}
%\footnote{$^{*}$These authors contributed equally to this work}
\affiliation{\vspace*{2mm}
%$^1$\textit{JSPS International Research Fellow, Graduate University of Science, Kyoto University, Japan}\vspace*{2mm}\\
%$^1$\textit{Graduate University of Science, Kyoto University, Japan}\vspace*{2mm}\\
$^1$\textit{Institute For Interdisciplinary Research in Science and Education},\\ 
\it{ICISE, Quy Nhon Nam 55131, Gia Lai, Vietnam\vspace*{2mm}\\}
%$^3$\textit{Institute of Physics, Vietnam Academy of Science and Technology, Hanoi 10000, Vietnam}, \vspace*{2mm}\\
$^2$\textit{Graduate University of Science and Technology, Vietnam Academy of Science and Technology, Hanoi 10000, Vietnam.}\\
}

\date{\today}% It is always \today, today,

\begin{abstract}
In this paper, we present an one of the most sensitive test of $CPT$ symmetry in the leptonic sector by probing the atmospheric neutrino mass-squared splittings, $\Delta m^2_{31}$ and $\Delta \overline{m}^2_{31}$, using simulated data from on-going reactor-based JUNO and future accelerator-based DUNE and Hyper-Kamiokande experiments. Our study focuses on the potential neutrino-antineutrino mass asymmetry $\delta_{\nu\overline{\nu}}(\Delta m^2_{31}) = \Delta m^2_{31} - \Delta \overline{m}^2_{31}$, achieving unprecedented precision by exploiting the high statistics and reduced systematic uncertainties of these facilities. The combined analysis yields a projected sensitivity to $CPT$ violation at the level of $\sim 1.7\times 10^{-5}~\text{eV}^2$ at $3\sigma$ confidence level. These results highlight the crucial role of multi-experiment synergies in testing fundamental symmetries of nature. The work also investigates impact of potential bias on the (anti-)neutrino mass-splitting measurement due to the neutrino interactions with accelerator-based neutrino experiments, and conclude that it must be improved by at least $39\%$ compared to T2K simulated-benchmark to achieve a sub-percent precision probe of $CPT$ violation.   
\end{abstract}

\maketitle

%\tableofcontents

%%%%%%%%%%%%%%%%%%%%%%%%%%%%%%%%%%%%%%%%%%%%%%%%%%%%%%%%%
%%%%%%%%%%%%%%%%%%%%%%%%%%%%%%%%%%%%%%%%%%%%%%%%%%%%%%%%%
\section{\label{intro} Introduction}
$CPT$ symmetry, the combined transformation of charge conjugation (C), parity (P), and time reversal (T), is a cornerstone of local, Lorentz-invariant quantum field theories \cite{Luders1954,1955nbdpbookPauli,Jost1957,Schwinger1958,johnbell}. Its potential violation would not only a signal of new physics beyond the Standard Model but also challenge our understanding of spacetime symmetries and quantum mechanics itself. Among the various sectors where $CPT$ violation can be probed, lepton sector offers a uniquely sensitive test due to the neutrino oscillation~\cite{Barger_2000,Murayama_2004, Barenboim2018, Barenboim2020, Barenboim_2001, Barenboim:2002hx, barenboim2023}, which is a macroscopic quantum interference phenomenon that depends on tiny mass-squared differences and other leptonic mixing parameters~\cite{maki1962remarks,pontecorvo1958mesonium}. Present landscape of neutrino oscillation~\cite{ParticleDataGroup:2024cfk} is driven by two scales: $\Delta m^2_{21}=m^2_2-m_1^2\approx 7.5\times 10^{-5}~\text{eV}^2$ or called solar sector and $|\Delta m^2_{31}=|m^3_2-m_1^2|\approx 2.5\times 10^{-3}~\text{eV}^2$ or atmospheric sector.

In a previous study~\cite{Ngoc:2022uhg}, we explored the possible $CPT$ violation in the atmospheric sector through both mass-squared splittings ($\Delta m^2_{31}$ and $\Delta \overline{m}^2_{31}$) and mixing angles ($\theta_{23}$ and $\overline{\theta}_{23}$), using simulated data from on-going accelerator-based T2K, NO$\nu$A, and branch-new reactor-based JUNO. The combined analysis offers a stringent constraints of $|\delta_{\nu\overline{\nu}}(\Delta m^2_{31})| < 5.3 \times 10^{-5}~\text{eV}^2$ and  $|\delta_{\nu\overline{\nu}}(\sin^2\theta_{23})| < 0.1$ at $3\sigma$ confidence level (CL) in case the data is found to be consisent with $CPT$ invariance. This work advances previous studies in three key aspects: (i) by focusing exclusively on the atmospheric mass-squared splitting as the most robust $CPT$-sensitive observable; (ii) by exploiting the synergy of branch-new JUNO experiment~\cite{djurcic2015juno} and under-construction Hyper-Kamiokande (Hyper-K) \cite{Abe:2018uyc} and DUNE~\cite{DUNE:2020lwj}; and (iii) by investigating the impact of the potential bias in the neutrino mass-squared splitting due to the neutrino interaction model. While the mixing angle is experimentally accessible for full phenomenology, its determination suffers from intrinsic octant degeneracies, and strong correlations with the $CP$-violating phase. These factors may introduce significant systematic uncertainties and hinder the clean extraction of $CPT$-violating signatures. By contrast, the atmospheric mass-squared splitting can be determined with high precision from $\nu_\mu$ and $\bar{\nu}_\mu$ \textit{disappearance} channels, with minimal dependence on the unknown $CP$ phase \cite{Nunokawa_2008, de_Salas_2021}. Therefore, it provides a more robust and theoretically transparent observable for $CPT$ tests. For this reason, our analysis in the present work focuses exclusively on the parameter pair $\Delta m^2_{31}$ and $\Delta \overline{m}^2_{31}$, and the observable asymmetry $\delta_{\nu\overline{\nu}}(\Delta m^2_{31})$ with assumption that other parameters are $CPT$-invarant. While previous sensitivity studies relied primarily on the current-generation long-baseline experiments T2K and NO$\nu$A, future facilities will deliver far superior precision of $\Delta m^2_{31}$ and $\Delta \overline{m}^2_{31}$ parameters. DUNE~\cite{DUNE:2020lwj}, with long baseline of 1300 km and high-resolution liquid-argon technology, will significantly enhance sensitivity to the neutrino mass ordering and provide ultra-precise measurements of $\Delta m^2_{31}$ and $\Delta \overline{m}^2_{31}$. Hyper-K~\cite{Abe:2018uyc}, with nearly an order of magnitude larger fiducial mass than Super-Kamiokande~\cite{fukuda2003super}, will collect enormous statistics in both neutrino and antineutrino modes. Its well-understood water Cherenkov technology and upgraded beam power offer percent-level precision on atmospheric parameters. JUNO~\cite{djurcic2015juno}, a reactor-based experiment with sub-percent energy resolution, will pin down the antineutrino mass-squared splitting ($\Delta \overline{m}^2_{31}$) with unrivaled precision, providing the necessary complement to accelerator-based measurements in testing $CPT$ symmetry. Together, these experiments offer the statistical power, systematic control, and complementary baselines required to probe $CPT$ violation with unprecedented sensitivity.
Our analysis demonstrates that these next-generation facilities, when combined, can push the sensitivity with $\delta_{\nu\overline{\nu}}(\Delta m^2_{31})$ to a factor of four improvement over the possible constraint set by T2K, NO$\nu$A, and JUNO~\cite{Ngoc:2022uhg}. This makes the atmospheric neutrino mass-splitting a promising channel for testing fundamental symmetries. 

The paper is structured as follows: Section~\ref{sec:experiment} describes the simulation framework and experimental configurations. Section~\ref{sec:current} presents current constraints on $CPT$ symmetry. Section~\ref{sec:CPTbound} investigates the bounds on the possible $CPT$ violation. Section~\ref{sec:CPTsensi} explores future sensitivity to $CPT$ violating parameter $\delta_{\nu\overline{\nu}}(\Delta m^2_{31})$, assesses the robustness of our finding under variations of the mixing angle $\theta_{23}$ and evaluate the impact due to the potential bias on $\Delta m^2_{31}$ with accelerator-based measurements. Finally, we summarize our findings and conclude in Section~\ref{sec:fin}.

%%%%%%%%%%%%%%%%%%%%%%%%%%%%%%%%%%%%%%%%%%%%%%%%%%%%%%%%%%%%%%%%%%%
%%%%%%%%%%%%%%%%%%%%%%%%%%%%%%%%%%%%%%%%%%%%%%%%%%%%%%%%%%%%%%%%%%%
\section{\label{sec:experiment} Methodology and Simulation framework}
%%%%%%%%%%%%%%%%%%%%%%%%%%%%%%%%%%%%%%%%%%%%%%%%%%%%%%%%%%%%%%%%%%%
%%%%%%%%%%%%%%%%%%%%%%%%%%%%%%%%%%%%%%%%%%%%%%%%%%%%%%%%%%%%%%%%%%%
The study of neutrino oscillations has entered an era of unprecedented precision, driven by a global network of accelerator-based and reactor-based experiments~\textcolor{red}{\cite{ParticleDataGroup:2024cfk}}. These facilities are pushing the boundaries of our understanding of neutrino properties, including mixing parameters, mass ordering, and potential violations of fundamental discrete symmetries like $CP$ and $CPT$. At the forefront of these efforts are the DUNE,  Hyper-K, and JUNO experiments, each offering unique capabilities to address open questions in neutrino physics.

The Jiangmen Underground Neutrino Observatory (JUNO)~\cite{djurcic2015juno} is a reactor-based experiment in China designed to measure the neutrino mass ordering and oscillation parameters with unprecedented precision by studying electron antineutrino \textit{disappearance} ($\overline{\nu}_e \rightarrow \overline{\nu}_e$). The detector consists of a 20-kton liquid scintillator, located 53 km from multiple nuclear reactors, optimized to observe interference between the solar and atmospheric mass splittings. This configuration provides exceptional energy resolution (3$\%$/$\sqrt{E}$) for precisely measuring $\Delta \overline{m}^2_{21}$ and \dmbaratm\ to better than 1$\%$ precision~\cite{juno_2022}. Of particular importance for $CPT$ tests, JUNO's precision measurement of the antineutrino \dmbaratm\ parameter will provide crucial independent constraints to compare with accelerator-based neutrino measurements. The experiment's ability to determine the neutrino mass ordering of greater than $3\sigma$ CL enhances its value for global neutrino physics analyses. JUNO commission happened in early of 2026 and its first result~\cite{abusleme2026measurement} based on just two months of operation already provide world-leading constraints on the solar neutrino paramters.

The Hyper-K~\cite{Abe:2018uyc} project represents a major leap forward in water Cherenkov detection technology. Scheduled to begin operations in 2028, Hyper-K will feature a detector fiducial volume of 187 kton (8.4 times larger than its predecessor Super-Kamiokande ~\cite{fukuda2003super}), coupled with an upgraded J-PARC beam capable of delivering 1.3 MW proton power. This enormous increase in statistics in both neutrino and antineutrino modes will significantly improve the sensitivity to $CP$ violation and neutrino mass ordering with the $\nu_e$ and $\overline{\nu}_e$ \textit{appearance} measurements. The sub-percent-level precision achievable on ($\Delta m^2_{31}$, $\Delta \overline{m}^2_{31}$) measurements in $\nu_\mu$ and $\overline{\nu}_\mu$ \textit{disappearance} channels \cite{hksensi.2505.15019} will enable stringent test of the $CPT$ invariance.

The Deep Underground Neutrino Experiment (DUNE)~\cite{DUNE:2020lwj}, currently under construction in the United States, represents a major advancement in long-baseline neutrino physics. DUNE will utilize a high-power (1.2 MW, upgradeable to 2.4 MW) neutrino beam produced at Fermilab and directed 1300 km toward the Sanford Underground Research Facility in South Dakota. The far detector will consist of four 10-kton modules of liquid argon time projection chambers (LArTPCs), enabling excellent tracking and calorimetry. DUNE’s long baseline enhances matter effects, improving sensitivity to the mass ordering and allowing precise measurements of neutrino and antineutrino oscillation parameters. The high-resolution LArTPC technology also allows a statistically robust separation of $\nu_\mu$ and $\overline{\nu}_\mu$ charged-current events through beam mode and interaction-level differences \cite{Kelly_2019, CHATTERJEE2024138838}, which is essential for probing $CPT$ violation in the atmospheric sector.

\begin{figure}
    \centering
    \includegraphics[width=0.95\linewidth]{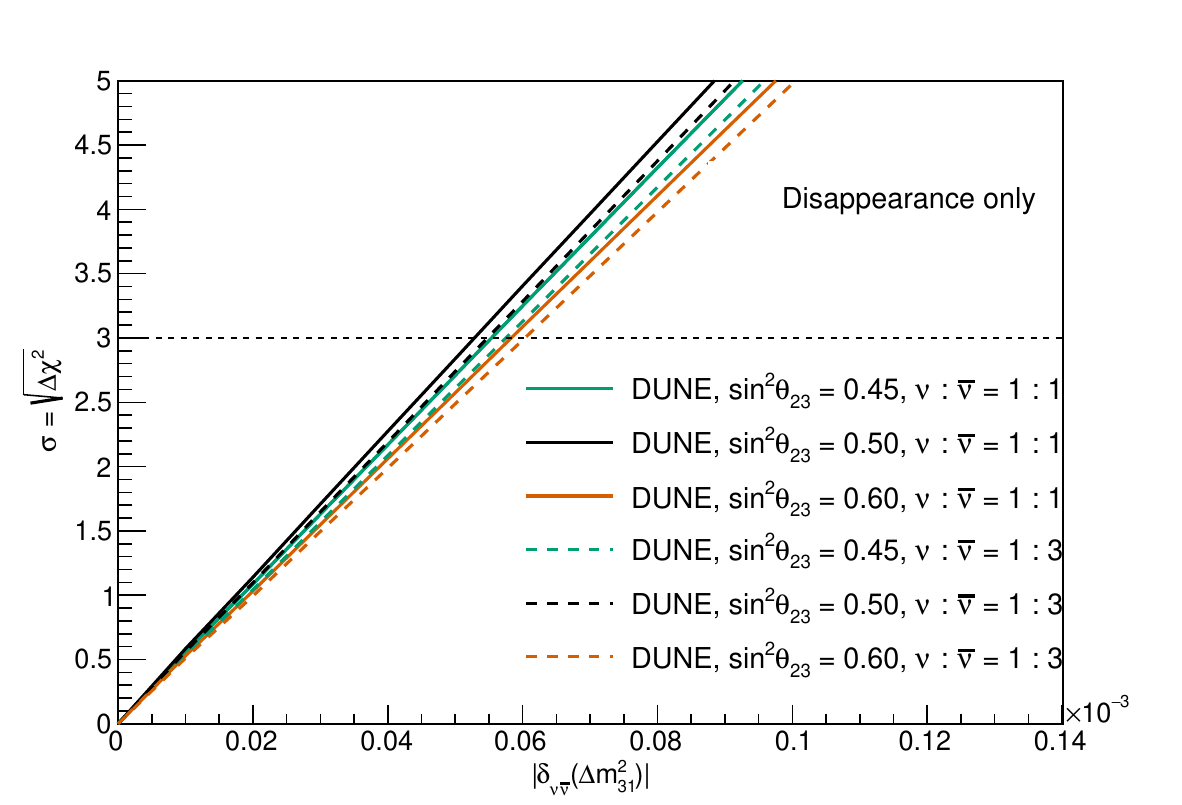}
    \includegraphics[width=0.95\linewidth]{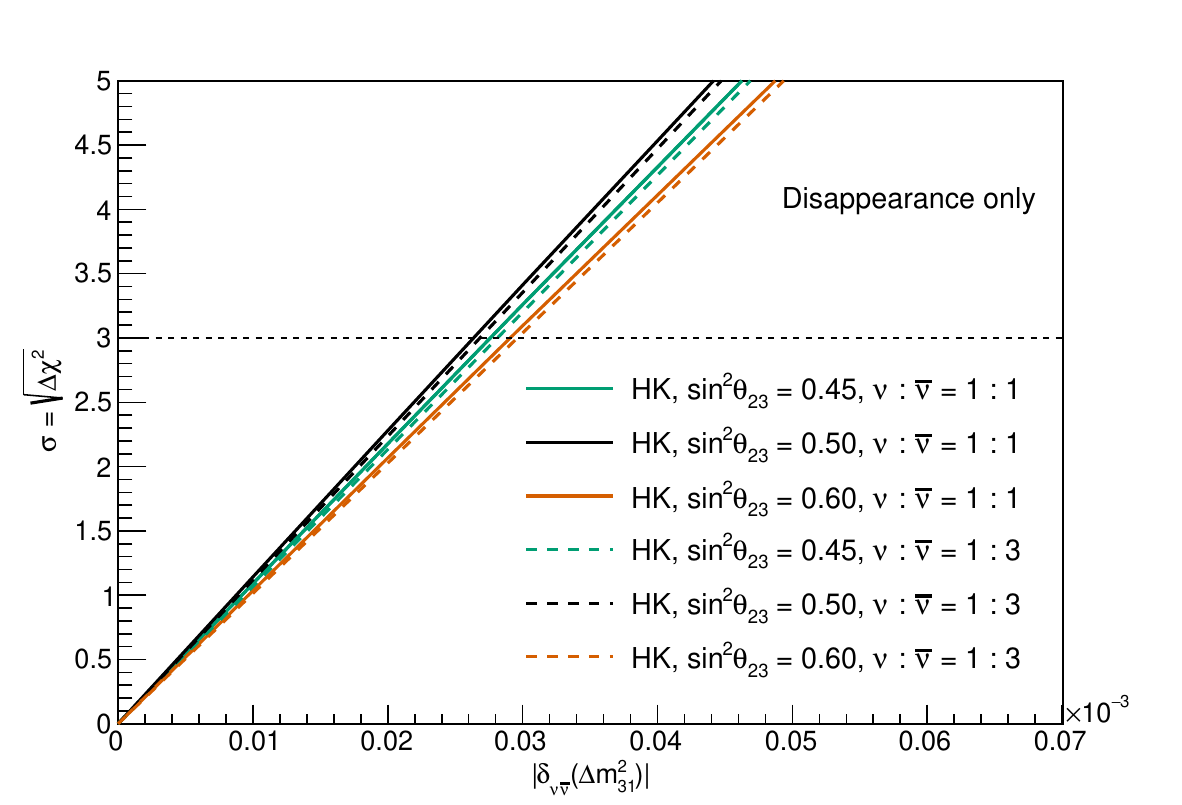}
    \caption{Comparison of statistical configurations for $CPT$ sensitivity in DUNE (top) and Hyper-K (bottom). For DUNE, the 1:1 neutrino–antineutrino running configuration improves the $CPT$ sensitivity by approximately 4\% compared to the 1:3 configuration, while for Hyper-K the corresponding improvement is about 2\%.}
    \label{fig:compareStat}
\end{figure}

Our analysis employs the extended GLoBES software package~\cite{Huber:2004ka,huber2007new} to calculate the $CPT$ test with key simulation parameters summarized in Table \ref{tab:simulation}. The configurations for different experiments are implemented as follows: The simulation for JUNO adopts the detailed setups from Ref.~\cite{Ngoc:2022uhg}, including detector specifications, beam fluxes, energy resolutions, and systematic uncertainties; DUNE's configuration follows closely the official design~\cite{duneglobes}; Hyper-K parameters are taken from the Technical Design Report~\cite{hktdr} which are also described in detail in Ref.~\cite{quyenprd2025}. We implement customized run plans of neutrino-to-antineutrino ratio 1:1 for both DUNE and Hyper-K. These ratios were optimized through sensitivity studies to maximize $CPT$-violation discovery potential. Specifically, for DUNE and Hyper-K, the 1:1 configuration performs $\sim 4\%$ and $\sim 2\%$ better than 1:3, respectively, as shown in Fig.~\ref{fig:compareStat}. The 1:1 ratio is motivated by the goal of maximizing sensitivity to $CPT$ violation, as it provides balanced statistics between neutrino and antineutrino channels. We emphasize that this is a physics-motivated alternative scenario, not a replacement for the official collaboration-optimized run plans.

For accelerator-based experiments, the $\nu_\mu$ and $\overline{\nu}_\mu$ \textit{disappearance} samples are used, while for reactor-based experiment JUNO, the $\overline{\nu}_e$ \textit{disappearance} data are used. All simulations include full three-flavor oscillation probability computational engine, abstract detector setups, with a GLoBES extension to handle independent neutrino and antineutrino oscillation parameters. Throughout the study, the \textit{normal} mass ordering is assumed to be true for both neutrinos and anti-neutrinos. The nominal values of the oscillation parameters and corresponding uncertainties are shown in Table \ref{tab:nuoscpara}, where the atmospheric mass-squared splittings ($\Delta m^2_{31},\ \Delta \overline{m}^2_{31}$) are derived from the measurements of T2K \cite{abe2021t2k} while the others are from the global fit \cite{deSalas2020}. 

For the core analysis presented in Sections \ref{sec:CPTbound} and \ref{sec:CPTsensi}, we employ the standard GLoBES Poisson log-likelihood $\chi^2$ with full systematic treatment: 
\begin{equation}
    \chi^2 = \chi^2_{stat} + \chi^2_{pull} + \chi^2_{prior},
\end{equation}
where $\chi^2_{stat}$ is the statistical term, computed using the standard Poisson likelihood implemented internally in GLoBES, $\chi^2_{pull}$ accounts for all systematic uncertainties implemented as pull parameters in the AEDL files, and $ \chi^2_{prior}$ represents external Gaussian priors on oscillation parameters and matter density. The predicted event rates are calculated for each experimental configuration based on the oscillation probabilities in the disappearance channels, which depend on the oscillation parameter $\Delta m^2_{31}$ for neutrinos and $\Delta \overline{m}^2_{31}$ for anti-neutrinos. The $\chi^2$ function is computed for each individual experiment and then minimized over all relevant oscillation parameters and nuisance parameters except $\Delta m^2_{31}$ and $\Delta \overline{m}^2_{31}$. The $\Delta \chi^2 = \chi^2 - \chi^2_{min}$ is defined and the statistical significance for excluding $CPT$ conservation is performed as $\sigma = \sqrt{\Delta \chi^2}$.

\begin{table*}[htbp]
    \centering
 \begin{tabular}{l|c|c|c|c|c}
    \hline
    \hline
         & JUNO & DUNE & Hyper-K  \\
    \hline
      Baseline (km)  & 52.5 & 1285 & 295 \\
      Matter density ($g/cm^3$) & 2.6 & 2.85 & 2.6 \\
      Detector mass (kt) & 20 & 40 & 187  \\
     Exposure (years) & 6 & 13 ($\sim 10$ years nominal)  & 10  \\
      & & ($\nu:\overline{\nu} = 6.5 : 6.5$) & ($\nu:\overline{\nu} = 5.0 : 5.0$) \\
      Power & 26.6 GWth & 1.2 MW & 1.3 MW \\ 
      Energy window & 1.8 - 9.0 MeV & 0.5 - 18.0 GeV & 0.1 - 5.05 GeV \\
     Number of bins & 200 & 80 & 99 \\
   Energy resolution & $2.9\%/\sqrt{E~(MeV)}$ & $15-20\%$ \footnote{via smearing matrix} &  $3.0\%/\sqrt{E~(GeV)}$\\ 
      \hline
    Signal detection efficiency & $73\% ~ \overline{\nu}_e$ & $> 90\%$\footnote{at flux peak for $\nu_\mu$ and $\overline{\nu}_\mu$ signals, implemented as post-smearing efficiencies in GLoBES (file Eff.inc)~\cite{duneglobes}}   &  ($69.7\%, 15.7\%$)$\nu_\mu$ and ($73.1\%, 41.5\%$) $\overline{\nu}_\mu$\footnote{(CCQE, CC-nonQE), for more details, please refer to Appendix A of Ref.~\cite{quyenprd2025}} \\
    \hline
    Signal sys. errors & $1.0\%$ ($0.5\%$)\footnote{normalization (calibration) error} & $5.0\%$ & $3.6\% ~(2.5\%)$  \\
      Background sys. errors &  &  $5.0\% \nu_e$, $5.0\% \nu_\mu$, $20.0\% \nu_\tau$, $5.0\% \overline{\nu}_e$, $5.0\%$ NC & $10.0\% ~(2.5\%)$ \\
    \hline
    \hline
    \end{tabular}
    \caption{Experimental configurations used for the analysis. We follow Ref.~\cite{Ngoc:2022uhg} for JUNO setup. The parameters of DUNE and Hyper-K are taken from Ref.~\cite{duneglobes} and Ref.~\cite{hktdr}, respectively. The neutrino-to-antineutrino ratios are set to 1:1 for DUNE and Hyper-K to optimize the sensitivity to $CPT$ violation.}
    \label{tab:simulation}
\end{table*}

\begin{table}
    \centering
    \begin{tabular}{l|c|c}
    \hline\hline
    Parameter & Central value & Error\\\hline
    %$\sin^{2}\theta_{23}$ & $0.51$\\
    %$\sin^{2}\overline{\theta}_{23}$ & $0.43$\\
    $\Delta m^{2}_{31} $ & $2.55\times 10^{-3}\text{eV}^{2}$ & $4\%$\\
    $\Delta \overline{m}^{2}_{31}$ & $2.61\times 10^{-3}\text{eV}^{2}$ & $7\%$\\
    & \\        
    $\sin^{2}\theta_{12},~ \sin^{2}\overline{\theta}_{12}$ & $0.318$ & $5\%$\\ 
    $\sin^{2}\theta_{13},~ \sin^{2}\overline{\theta}_{13}$ & $0.022$ & $3\%$\\ 
    $\sin^{2}\theta_{23},~ \sin^{2}\overline{\theta}_{23}$ & $0.574$ & $4\%$\\
    $\delta_{\text{CP}},~ \overline{\delta}_{\text{CP}}$ & $1.08\pi$~ rad & $12\%$\\
    $\Delta m^{2}_{21},~ \Delta \overline{m}^{2}_{21}$ & $7.50\times10^{-5}\text{eV}^{2}$ &$3\%$\\
    \hline
    \end{tabular}
    \caption{Values and corresponding uncertainties of nominal parameters used in our study. $\Delta m^{2}_{31}$, $\Delta \overline{m}^{2}_{31}$ are derived from the recent T2K measurements\cite{t2kprd2023update} while the rests are from the global analysis~\cite{deSalas2020}. Normal mass ordering is assumed.}
    \label{tab:nuoscpara}
\end{table}
%%%%%%%%%%%%%%%%%%%%%%%%%%%%%%%%%%%%%%%%%%%%%%%%%%%%%%%%%%%%%%%%%%%
%%%%%%%%%%%%%%%%%%%%%%%%%%%%%%%%%%%%%%%%%%%%%%%%%%%%%%%%%%%%%%%%%%%
\section{\label{sec:current} Current constraints on $CPT$ violation from T2K, NO$\nu$A, and Daya Bay}
%%%%%%%%%%%%%%%%%%%%%%%%%%%%%%%%%%%%%%%%%%%%%%%%%%%%%%%%%%%%%%%%%%%
%%%%%%%%%%%%%%%%%%%%%%%%%%%%%%%%%%%%%%%%%%%%%%%%%%%%%%%%%%%%%%%%%%%
In this section, we present the status of $CPT$ symmetry tests using neutrino oscillation data from T2K and NO$\nu$A, combined with Daya Bay's constraint on the antineutrino mass-squared splitting $\Delta \overline{m}^2_{31}$. While T2K and NO$\nu$A provide sensitive measurements of both neutrino $\Delta m^2_{32}$ and antineutrino $\Delta \overline{m}^2_{32}$ parameters, Daya Bay's high-precision reactor antineutrino data further constrains $\Delta \overline{m}^2_{32}$. For consistency with the default GLoBES parameterization and standard
oscillation analyses, we express the atmospheric mass-squared splittings in terms of $\Delta m^2_{31}$ ($\Delta \overline{m}^2_{31}$), rather than $\Delta m^2_{32}$
($\Delta \overline{m}^2_{32}$). This choice facilitates a direct implementation within the modified GLoBES simulation framework and allows for a transparent comparison with existing
literature. Under the assumption of normal mass ordering, the conversion is given by $\Delta m^2_{31} = \Delta m^2_{32} + \Delta m^2_{21}$ for neutrinos and
$\Delta \overline{m}^2_{31} = \Delta \overline{m}^2_{32} + \Delta \overline{m}^2_{21}$ for antineutrinos. The uncertainties are also converted accordingly ($\sigma_{31} = \sqrt{\sigma^2_{32} + \sigma^2_{21}}$). Since $\Delta m^2_{21}$ scale is significantly smaller (around 3\%) than $\Delta m^2_{31}$ and known with high 1\% precision, it is found that a common $\Delta m^2_{21}=\Delta \overline{m}^2_{21}$ assumption does not affect the $CPT$-violating observable $\delta_{\nu\bar{\nu}}(\Delta m^2_{31})$.

In 2023, the T2K collaboration's updated analysis of muon neutrino and antineutrino disappearance, utilizing a combined exposure of $3.6\times 10^{21}$ protons on target (POTs), provides precise measurements of the atmospheric neutrino oscillation parameters \cite{t2kprd2023update}. By fitting simultaneously neutrino and antineutrino data with independent set of oscillation parameters, the study yields best-fit values of $\Delta m^2_{32} = 2.48^{+0.05}_{-0.06} \times 10^{-3} \text{eV}^2$ for neutrino and  $\Delta \overline{m}^2_{32} = 2.53^{+0.10}_{-0.11} \times 10^{-3} \text{eV}^2$ for antineutrino, assuming normal mass ordering. These results are derived from a fit to the reconstructed energy and angular distributions of $1R\mu$ events at Super-Kamiokande, constrained by near-detector measurements that significantly reduce systematic uncertainties related to flux and cross-section modeling. The consistency between neutrino and antineutrino parameters supports the standard three-flavor oscillation framework. The analysis benefits from improved hadron production data from NA61/SHINE, refined neutrino-nucleon/nucleus interaction models, and an expanded dataset, leading to a reduction in the total systematic uncertainty by approximately 45\% for the neutrino mode compared to previous T2K results. 

In a dedicated analysis of muon neutrino and antineutrino \textit{disappearance} data collected by the NO$\nu$A experiment \cite{nova20192d118E20POT}, measurements of the atmospheric mass-squared differences were obtained without assuming $CPT$ invariance. Using exposures of $8.85\times 10^{20}$ POTs in neutrino mode and $12.33\times10^{20}$ POTS in antineutrino mode, the analysis yielded best-fit values of $\Delta m^2_{32} = 2.48^{+0.07}_{-0.09} \times 10^{-3} \text{eV}^2$ for neutrino and  $\Delta \overline{m}^2_{32} = 2.55^{+0.12}_{-0.13} \times 10^{-3} \text{eV}^2$ for antineutrino.

The Daya Bay reactor neutrino experiment has performed a precision measurement of the atmospheric mass-squared difference $\Delta \overline{m}^2_{32}$ using a dataset spanning 3158 days of operation \cite{dayabayprd2023}. By analyzing over 5.55 million inverse beta-decay (IBD) events with neutron capture on gadolinium, and employing improved energy calibration and background suppression techniques, the collaboration extracted oscillation parameters through a relative comparison of antineutrino energy spectra at multiple baselines. The resulting best-fit value for the atmospheric mass-squared difference is $\Delta \overline{m}^2_{32} = 2.466^{+0.06}_{-0.06} \times 10^{-3} \text{eV}^2$ for normal mass ordering. This result demonstrates excellent consistency with measurements from accelerator-based experiments, reinforcing the robustness of the three-flavor oscillation framework and providing a good constraint on the antineutrino parameter for the efforts of testing $CPT$.

\begin{figure}
\includegraphics[width=0.45\textwidth]{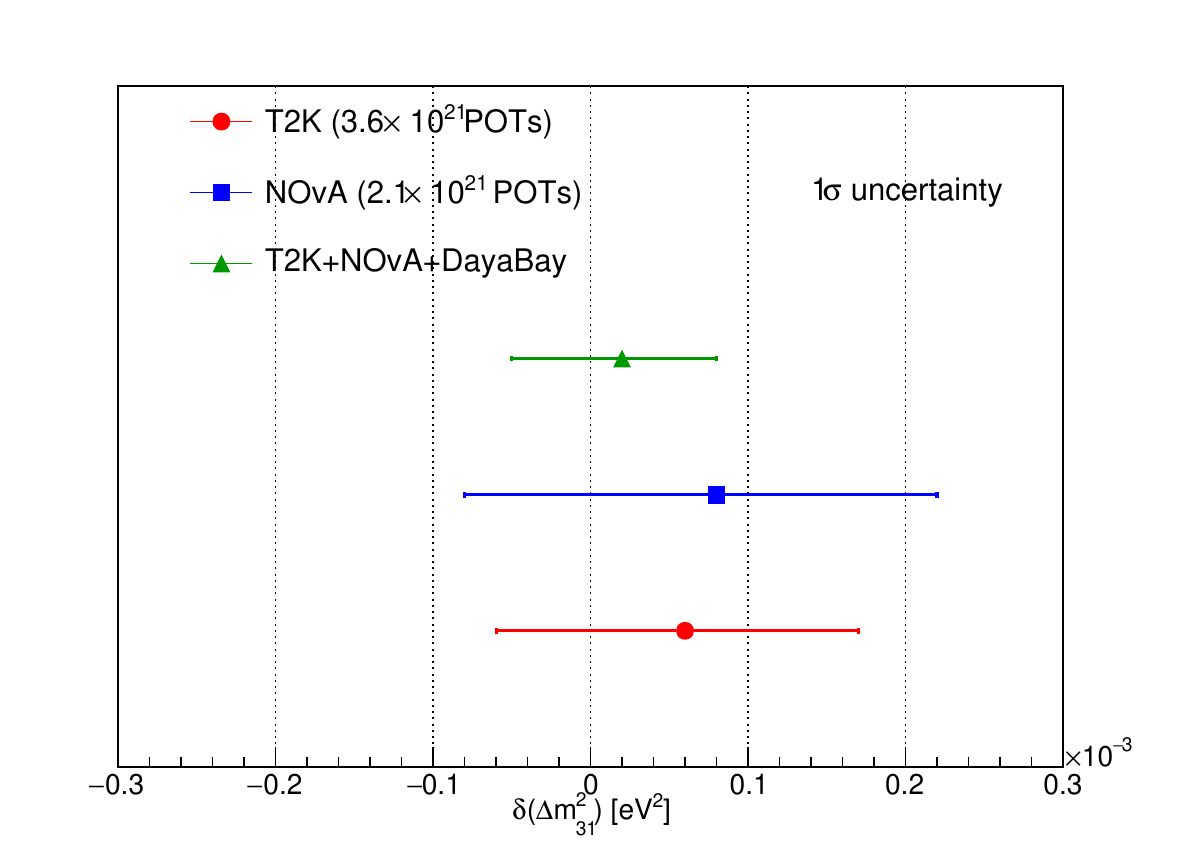}
\caption{\label{fig:latestt2knovadayabay} The current analyses from T2K \cite{t2kprd2023update} (red) and NO$\nu$A \cite{nova20192d118E20POT} (blue), along with their combined results incorporating Daya Bay \cite{dayabayprd2023} (green) data within 1$\sigma$ uncertainty.}
\end{figure}

 For the combined reinterpretation, we define the $\chi^2$ function as follows
\begin{equation}\label{eq:chi2func}
    \chi^2=\sum_{i}\frac{[(\Delta m^2_{31})_{test} - (\Delta m^2_{31})_i]^2}{\sigma(\Delta m^2_{31})^2_i},
\end{equation}
where the index $i$ runs over the T2K, NO$\nu$A, and Daya Bay, each contributing their best-fit values and associated uncertainties. The global best-fit point corresponds to the combined value of $\Delta m^2_{31} (\Delta \overline{m}^2_{31})$ when the $\chi^2$ function gets minimum. The associated uncertainties are evaluated at the $1\sigma$ confidence level using
\begin{equation}\label{eq:chi2func}
    \sigma(C.L.) = \sqrt{\Delta \chi^2} = \sqrt{\chi^2 - \chi^2_{min}}.
\end{equation}
The summary is provided in Table~\ref{tab:latestt2knovadayabay}, and the corresponding constraints are visualized in Fig.~\ref{fig:latestt2knovadayabay}. The results show no statistically significant indication of $CPT$ violation in the atmospheric sector, based on the current data from T2K and NO$\nu$A, as well as their combined interpretation including Daya Bay measurements.
\begin{table}
    \centering
    \begin{tabular}{l|c|c|c|c}
    \hline
    \hline
     NO is assumed & T2K & \nova & Daya Bay & Combined  \\
    \hline
      $\Delta m^2_{31}/10^{-3} \text{eV}^2$ & $2.55^{+0.05}_{-0.06}$ & $2.55^{+0.07}_{-0.09}$ & & $2.55^{+0.04}_{-0.05}$  \\
      $\Delta \overline{m}^2_{31}/10^{-3} \text{eV}^2$ & $2.61^{+0.10}_{-0.11}$ & $2.63^{+0.12}_{-0.13}$ & $2.54^{+0.06}_{-0.06}$ & $2.57^{+0.05}_{-0.05}$  \\
     $\sigma(\Delta m^2_{31})/10^{-3} \text{eV}^2$ & $0.06^{+0.11}_{-0.12}$ & $0.08^{+0.14}_{-0.16}$ & & $0.02^{+0.06}_{-0.07}$ \\
      \hline
    \end{tabular}
    \caption{Latest results on measurement of mass squared splittings $\Delta m^2_{31}$ and $\Delta \overline{m}^2_{31}$ from T2K \cite{t2kprd2023update}, NO$\nu$A \cite{nova20192d118E20POT}, and Daya Bay \cite{dayabayprd2023}.}
    \label{tab:latestt2knovadayabay}
\end{table}

%%%%%%%%%%%%%%%%%%%%%%%%%%%%%%%%%%%%%%%%%%%%%%%%%%%%%%%%%%%%%%%%%%%
%%%%%%%%%%%%%%%%%%%%%%%%%%%%%%%%%%%%%%%%%%%%%%%%%%%%%%%%%%%%%%%%%%%
\section{\label{sec:CPTbound} Stringent bounds on the neutrino-antineutrino mass asymmetry}
%\section{\label{sec:CPTbound} Stringent bounds on the \sout{possible $CPT$ violation} \textcolor{red}{neutrino-antineutrino mass asymmetry}}
%%%%%%%%%%%%%%%%%%%%%%%%%%%%%%%%%%%%%%%%%%%%%%%%%%%%%%%%%%%%%%%%%%%
%%%%%%%%%%%%%%%%%%%%%%%%%%%%%%%%%%%%%%%%%%%%%%%%%%%%%%%%%%%%%%%%%%%
In this section, we evaluate the expected bound of the neutrino-antineutrino mass asymmetry $\delta_{\nu\bar{\nu}}(\Delta m^2_{31})$ through a series of progressively constrained experimental configurations. For comparison, we first include the combined analysis of T2K, NO$\nu$A, and JUNO data, as reported in Ref.~\cite{Ngoc:2022uhg}. We then extend the study to next-generation experiments, considering the standalone capabilities of DUNE and Hyper-K, followed by their combinations with JUNO (DUNE+JUNO and Hyper-K+JUNO), and ultimately the full synergy of all three experiments (Hyper-K+DUNE+JUNO). Our analysis framework adopts exact $CPT$ conservation ($\delta_{\nu\overline{\nu}}(\Delta m^2_{31}) =0$) as the null hypothesis, with the true oscillation parameters fixed to T2K's most recent normal ordering measurements \cite{t2kprd2023update} ($\Delta {m}^2_{31}=\Delta\overline{m}^2_{31} = 2.55\times 10^{-3}\ \text{eV}^2$). The leptons and anti-lepton mixing matrices, which are normally parameterized by three mixing angles and one irreducible $CP$-violation phase, are assumed to be identical. For $\sin^2\theta_{23}$ and $\sin^2\overline{\theta}_{23}$, we put $\sin^2\theta_{23} = \sin^2\overline{\theta}_{23}$ and perform a dense scan across the currently allowed $3\sigma$ interval of [0.40, 0.62] using 23 evenly spaced values with a step size of 0.01. These choices cover the full experimentally allowed region and capture the key features of octant degeneracy and maximal mixing. The statistical methodology employs a frequentist $\chi^2$ approach to construct high-resolution two-dimensional parameter grids $(\Delta {m}^2_{31},\Delta\overline{m}^2_{31})$. The derived $\Delta\chi^2 = \chi^2-\chi^2_{min} $ profiles are interpreted through standard confidence levels, with particular attention to the 3$\sigma$ threshold as the benchmark for potential $CPT$ violation evidence. For accelerator-based Hyper-K and DUNE experiments, systematic error are implemented by normalization uncertainties in the signal and background event rates for all simulated samples, which effectively take into account uncertainties in under standing neutrino flux, interaction model, and detector response. No correlation is assumed among these parameters for simplicity. Signal section efficiency and energy resolution are also embedded in the simulation framework. For JUNO, the normalization factors for reactor neutrino flux, detector mass, signal spectral, calibrated energy scale, and reconstructed energy shape are implemented. The combined experimental configurations offer significantly enhanced sensitivity through three key mechanisms:
(i)  the three detection technologies - water Cherenkov (Hyper-K), liquid argon (DUNE), and liquid scintillator (JUNO) - carry largely uncorrelated systematic uncertainties, so a joint fit is expected to be less susceptible than any single experiment to technology-specific biases in cross-section modeling, energy reconstruction, and detector response; (ii) the ensemble probes oscillations over a broad range of oscillation wavelength, L/E, from reactor antineutrinos near the inverse-beta-decay threshold ($\sim 1.8$ MeV) in JUNO, through the sub-GeV to multi-GeV accelerator beams of Hyper-K and DUNE, to multi-GeV atmospheric neutrinos, thereby sampling over a wide range of interactions including the quasi-elastic, resonance, and deep-inelastic regimes; and (iii) electron anti-neutrino \emph{disappearance} in JUNO will provide a sub-percent precision for antineutrino sector parameters $\Delta \overline{m}^2_{21}, \Delta \overline{m}^2_{31}, \overline{\theta}_{12}
$ while muon (anti-) neutrino \emph{disappearance} in Hyper-K and DUNE will provide $\sim 0.5$\% precision measurements of $\Delta {m}^2_{31}, \Delta \overline{m}^2_{31}, {\theta}_{23}, \overline{\theta}_{23}
$ \cite{HyperK2025sensi, Abi_2020}; their combination help to resolve the correlations among the fitted parameters and  provide among the most sensitive measurement on the potential $CPT$ violation in the lepton sector and pave the way for further test for future.

\begin{figure}
\centering
\includegraphics[width=0.45\textwidth]{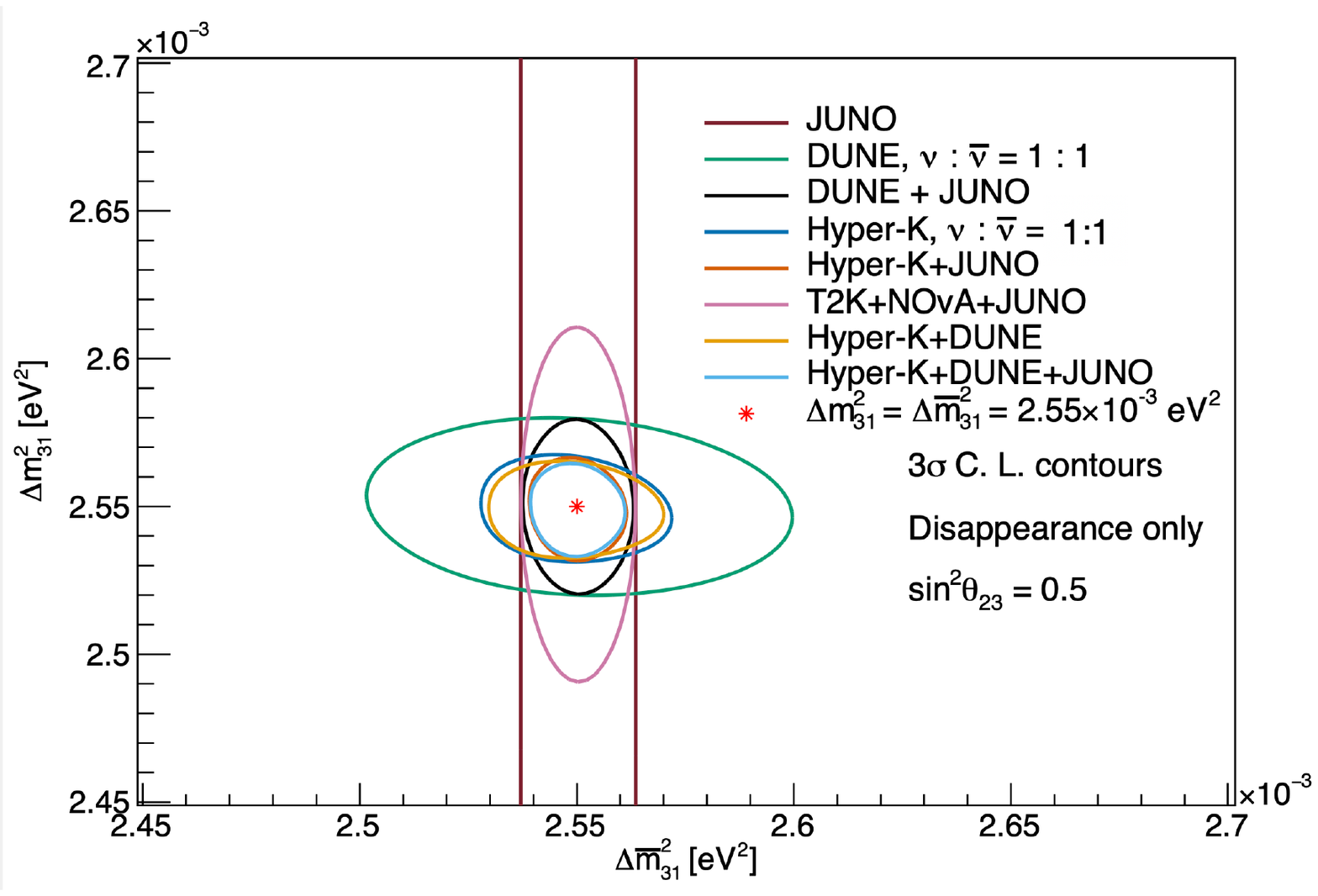}
\caption{\label{fig:contour2D} The two-dimensional contours of $\Delta m^2_{31}$ and $\Delta \overline{m}^2_{31}$ at 3$\sigma$ C. L. constraints on the $CPT$ violation from current and future neutrino oscillation experiments (combined T2K+NO$\nu$A+JUNO, DUNE, Hyper-K, JUNO, and their combinations).}
\end{figure}

\begin{figure}
\includegraphics[width=0.45\textwidth]{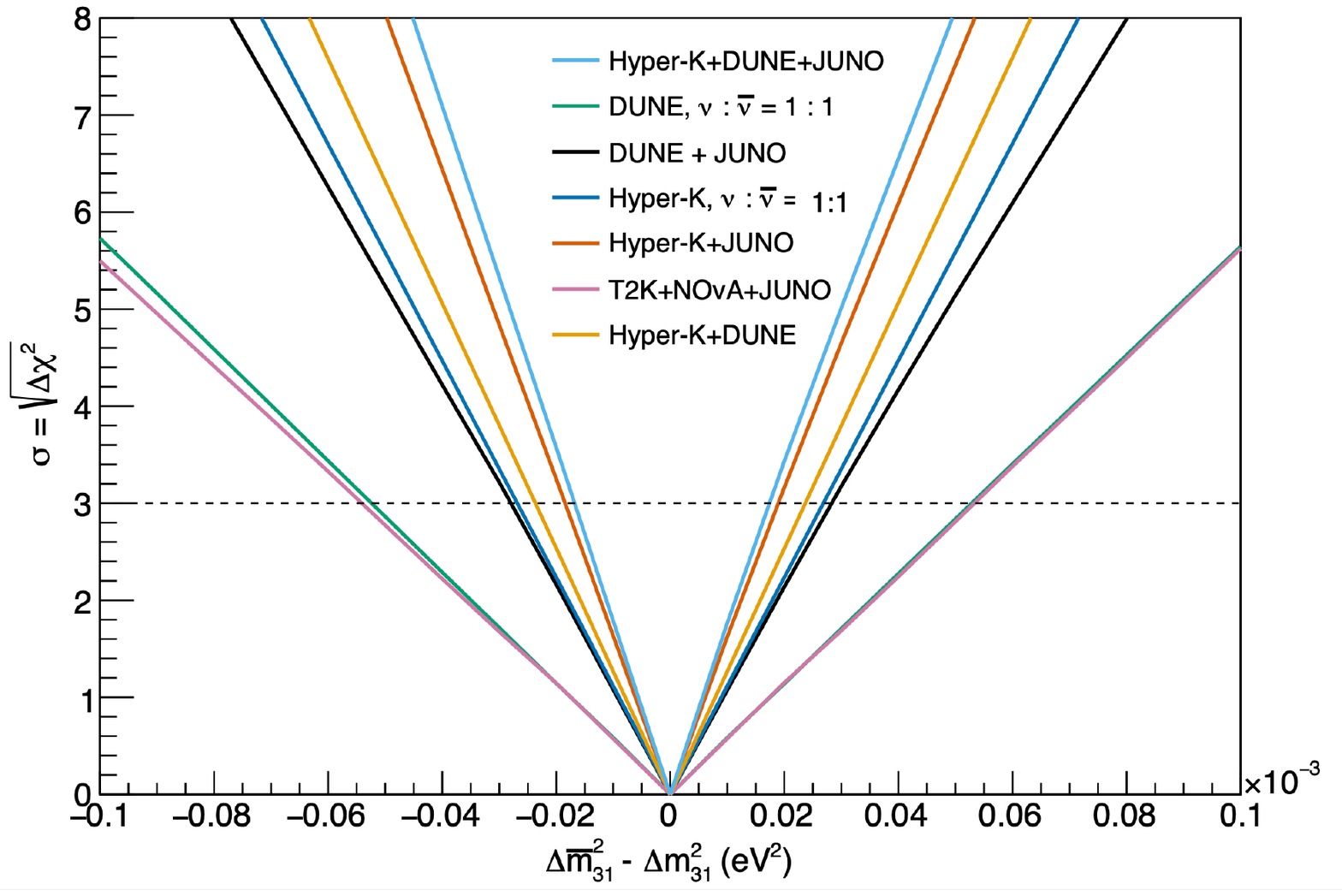}
\caption{\label{fig:contour1D} The bounds on possible $CPT$ violation with \dcptdm under various experimental configurations.}
\end{figure}

Table~\ref{tab:cptbound} summarizes the 3$\sigma$ C. L. upper limits on the true value of $|\delta_{\nu\overline{\nu}}(\Delta m^2_{31})|$ required to be consistent with the $CPT$ conservation hypothesis ($\delta_{\nu\overline{\nu}}(\Delta m^2_{31}) =0$) for various experimental configurations. The limits are calculated assuming $\sin^2\theta_{23} = \sin^2\overline{\theta}_{23}$, with three benchmark scenarios: lower octant ($\sin^2\theta_{23} = 0.45$), maximal mixing ($\sin^2\theta_{23} = 0.50$), and higher octant ($\sin^2\theta_{23} = 0.60$). Fig.~\ref{fig:contour2D} shows two-dimensional $\Delta m^2_{31}$ - $\Delta \overline{m}^2_{31}$ confidence contours, while Fig.~\ref{fig:contour1D} displays the corresponding one-dimensional $|\delta_{\nu\overline{\nu}}(\Delta m^2_{31})|$ parameter projections. These figures demonstrate the sensitivity of our analysis to potential $CPT$ violation effects under the assumption of exact maximal mixing.
The results reveal that current generation experiments (T2K and NO$\nu$A) combined with JUNO achieve sensitivities of $\sim 5\times 10^{-5}~\text{eV}^2$ across all considered $\theta_{23}$ values, which is comparable to the standalone DUNE performance. This consistency demonstrates the complementary roles of accelerator-based (T2K and NO$\nu$A) and reactor-based (JUNO) experiments in constraining $CPT$ violating effects. In Ref.~\cite{Barenboim2018}, a limit of $|\delta_{\nu\overline{\nu}}(\Delta m^2_{31})| < 8.1\times 10^{-5},\text{eV}^2$ at 3$\sigma$ CL was reported. In contrast, our results yield a significantly improved limit of $5.4\times 10^{-5}\text{eV}^2$ at maximal mixing, as shown in Table~\ref{tab:cptbound}. This improvement can be primarily attributed to the updated experimental configuration adopted in our analysis, based on the GLoBES implementation provided by the DUNE collaboration \cite{duneglobes}. In particular, we assume a larger 40 kt liquid argon detector compared to 10 kt in Ref.~\cite{Barenboim2018}, a longer exposure of 6.5 years in both neutrino and antineutrino modes compared to 3.5+3.5 years, and a higher beam power of 1.2 MW. These upgrades lead to a substantial increase in statistics which directly translate into stronger constraints on $CPT$ sensitivity. DUNE shows slightly better performance than combined T2K+NO$\nu$A+JUNO. This reflects the fact that, although the addition of JUNO improves the overall constraint by reducing parameter degeneracies, the sensitivity remains limited by the lower statistics of T2K and NO$\nu$A.

\begin{figure}
\includegraphics[width=0.45\textwidth]{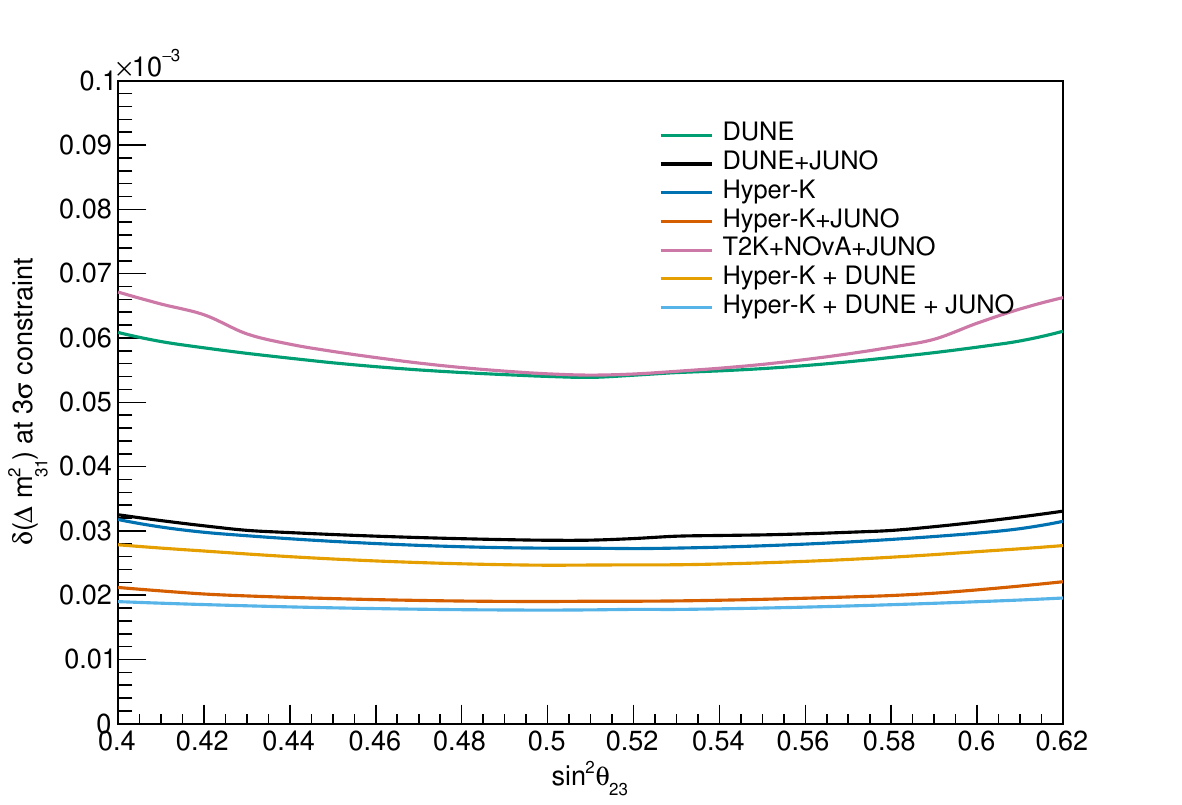}
\caption{\label{fig:bound3sigmath23} The bounds at 3$\sigma$ CL of \dcptdm across the 3$\sigma$ experimentally allowed range of $\sin^2\theta_{23} \in [0.40 - 0.62]$.}
\end{figure}

Hyper-K significantly improves the sensitivity ($\sim 2.7 - 2.9 \times 10^{-5}~\text{eV}^2$), surpassing T2K+NO$\nu$A+JUNO and DUNE by approximately a factor of two, and slightly outperforming DUNE+JUNO due to its huge detector volume and higher beam power. The most stringent constraints $\sim 2 \times 10^{-5}~\text{eV}^2$ emerge from Hyper-K combined with DUNE and/or JUNO, representing a $\sim 60\%$ improvement over T2K+NO$\nu$A+JUNO. Notably, maximal mixing ($\sin^2\theta_{23} = 0.5$) consistently gives the best sensitivity for all configurations, while deviations to non-maximal value degrade precision by $\sim 3 - 13 \%$, reflecting the impact of $\theta_{23}$ octant degeneracies on $CPT$ violation searches. The degradation becomes more significant as $\sin^2\theta_{23}$ moves further from 0.5 in either octant direction, as clearly demonstrated in Fig.~\ref{fig:bound3sigmath23}. The figure displays the 3$\sigma$ CL constraints on $|\delta_{\nu\overline{\nu}}(\Delta m^2_{31})|$ across the full 3$\sigma$ allowed range of $\sin^2\theta_{23}$ [0.40 - 0.62]. The parabolic shape of the sensitivity curves reflect the fundamental role of $\theta_{23}$ degeneracies in neutrino oscillation analyses, when $\theta_{23}$ deviates from maximal mixing, the increased correlation between $\theta_{23}$ and $\Delta m^2_{31}$ measurements leads to broader confidence intervals. The flatter sensitivity curves of DUNE, Hyper-K, and their combinations with JUNO demonstrate their abilities to resolve the $\theta_{23}$ octant degeneracy and mitigate its correlation with
$\Delta m^2_{31}$. This reduced parametric dependence results from their complementary neutrino and antineutrino beam spectra and enhanced statistical precision. Table \ref{tab:cptboundultimate} presents the expected upper limits on the magnitude of neutrino-antineutrino mass splitting difference $|\delta_{\nu\overline{\nu}}(\Delta m^2_{31})|$, that can be constrained at 3$\sigma$ CL under the assumption of $CPT$ conservation. The values are calculated assuming $\sin^2\theta_{23} = \sin^2\overline{\theta}_{23}$ and cover the 3$\sigma$ allowed range of $\sin^2\theta_{23}$ [0.40 - 0.62]. Among individual experimental configurations, Hyper-K demonstrates superior sensitivity to $CPT$ violating parameters, establishing the most rigorous limit of $3.09\times 10^{-5}~\text{eV}^2$ on $\delta_{\nu\overline{\nu}}(\Delta m^2_{31})$. This constraint can be further improved to $1.83\times 10^{-5}~\text{eV}^2$ through the Hyper-K+JUNO combined analysis and to $1.77\times 10^{-5}~\text{eV}^2$ if DUNE data is added, demonstrating the power of multi-experiment synergies in testing fundamental symmetries.

It is also interesting to compare our results with recent studies of $CPT$ violation in the solar sector, such as Ref.~\cite{barenboim2023}. An updated global analysis of current solar and KamLAND data yields $3\sigma$ CL. of $|\delta(\Delta m^2_{21})| < 3.7 \times 10^{-5}~\text{eV}^2$. The study in \cite{barenboim2023} projects that the combination of JUNO, Hyper-K, and DUNE could improve these limits by up to an order of magnitude, reaching a sensitivity of $|\delta(\Delta m^2_{21})| < 0.8 \times 10^{-5}~\text{eV}^2$ in an optimal configuration. Although that analysis focuses on the solar oscillation parameters, a similar trend is observed: the combination of on-going JUNO with next-generation experiments DUNE and Hyper-K, leads to a substantial improvement in sensitivity. While a direct comparison between the atmospheric and solar sectors is not straightforward due to the different energy scales and oscillation channels involved, it is noteworthy to know that the sensitivity we find for the atmospheric mass splitting ($|\delta_{\nu\overline{\nu}}(\Delta m^2_{31})| \sim 1.7\times 10^{-5}~\text{eV}^2$ with combined Hyper-K, DUNE, and JUNO) is of the same order of magnitude as the projected sensitivity for the solar mass splitting in the optimal scenario. This consistency underscores the remarkable potential of the next-generation neutrino facility network (DUNE, Hyper-K, and JUNO) to probe $CPT$ invariance across both mass scales with unprecedented precision, reaching the $\mathcal{O}(10^{-5}~\text{eV}^2)$ level.

\begin{table*}[htbp]
    \centering
    \begin{tabular}{c|c|c|c|c|c|c|c}
    \hline
    \hline
     $\sin^2\theta_{23} = \sin^2\overline{\theta}_{23}$ & T2K+NO$\nu$A+JUNO & DUNE  & DUNE+JUNO  & HK & HK+DUNE &HK+JUNO & HK+DUNE+JUNO \\
      &  [$eV^2$]& [$eV^2$] & [$eV^2$] &  [$eV^2$]& [$eV^2$] &  [$eV^2$]&  [$eV^2$]\\
    \hline
      0.45 & $5.79\times 10^{-5}$ & $5.61\times 10^{-5}$ & $2.94\times 10^{-5}$ & $2.80\times 10^{-5}$ & $2.55\times 10^{-5}$ & $1.69\times 10^{-5}$ & $1.63\times 10^{-5}$\\
      0.50 & $5.44\times 10^{-5}$ & $5.40\times 10^{-5}$ & $2.85\times 10^{-5}$ & $2.70\times 10^{-5}$ & $2.46\times 10^{-5}$ & $1.67\times 10^{-5}$ & $1.61\times 10^{-5}$ \\
      0.60 & $6.23\times 10^{-5}$ & $5.86\times 10^{-5}$ & $3.14\times 10^{-5}$ & $2.92\times 10^{-5} $& $2.67\times 10^{-5}$ & $1.78\times 10^{-5}$ & $1.72\times 10^{-5}$\\
      \hline
    \end{tabular}
    \caption{The upper limits at 3$\sigma$ CL on the $CPT$ violation with atmospheric mass-squared splittings for various experimental configurations. Assuming $\sin^2\theta_{23} = \sin^2\overline{\theta}_{23}$, values are calculated for three benchmark mixing scenarios (lower octant: 0.45, maximal mixing: 0.50, and higher octant: 0.60).}
    \label{tab:cptbound}
\end{table*}

\begin{table*}[htbp]
    \centering
    \begin{tabular}{c|c|c|c|c|c|c|c}
    \hline
    \hline
     $\sin^2\theta_{23} = \sin^2\overline{\theta}_{23}$ & T2K+NO$\nu$A+JUNO & DUNE & DUNE+JUNO & HK & HK+DUNE & HK+JUNO & HK+DUNE+JUNO \\
       &  [$eV^2$]& [$eV^2$] & [$eV^2$] &  [$eV^2$]& [$eV^2$] &  [$eV^2$]&  [$eV^2$]\\
    \hline
     0.40 - 0.62  & $6.71\times 10^{-5}$ & $6.10\times 10^{-5} $ & $3.31\times 10^{-5} $ & $3.09\times 10^{-5} $ & $2.77\times 10^{-5} $ & $1.83\times 10^{-5}$ & $1.77\times 10^{-5}$ \\
      \hline
    \end{tabular}
    \caption{The maximum values of upper limits at 3$\sigma$ CL on the $CPT$ violation with atmospheric mass-squared splittings for various experimental configurations obtained by scanning over the 3$\sigma$ allowed range of $\sin^2\theta_{23} = \sin^2\overline{\theta}_{23} \in [0.40 - 0.62]$.}
    \label{tab:cptboundultimate}
\end{table*}


%%%%%%%%%%%%%%%%%%%%%%%%%%%%%%%%%%%%%%%%%%%%%%%%%%%%%%%%%%%%%%%%%%%
%%%%%%%%%%%%%%%%%%%%%%%%%%%%%%%%%%%%%%%%%%%%%%%%%%%%%%%%%%%%%%%%%%%
\section{\label{sec:CPTsensi} The expected $CPT$ violation sensitivity}
%%%%%%%%%%%%%%%%%%%%%%%%%%%%%%%%%%%%%%%%%%%%%%%%%%%%%%%%%%%%%%%%%%%
%%%%%%%%%%%%%%%%%%%%%%%%%%%%%%%%%%%%%%%%%%%%%%%%%%%%%%%%%%%%%%%%%%%
\begin{figure}
\includegraphics[width=0.45\textwidth]{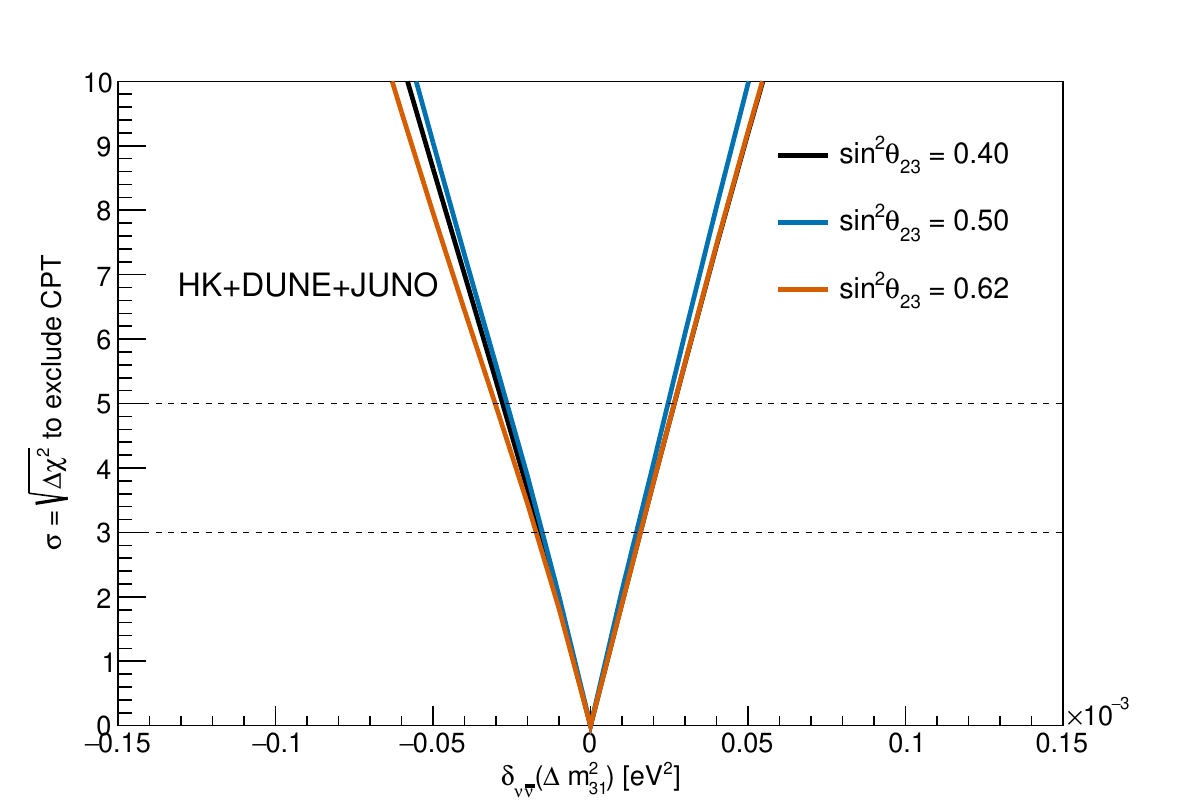}
\caption{\label{fig:cptsensidunehkjuno} The sensitivities to $CPT$ violation as a function of \textit{true} \dcptdm for Hyper-K+DUNE+JUNO at three different octant scenarios of $\sin^2\theta_{23} = \sin^2\overline{\theta}_{23}$: 0.40 (black), 0.50 (blue), and 0.62 (vermilion). $\Delta m^2_{31} 2.55\times 10^{-3}~\text{eV}^2$ is assumed to be \textit{true}.}
\end{figure}

\begin{figure}
\includegraphics[width=0.45\textwidth]{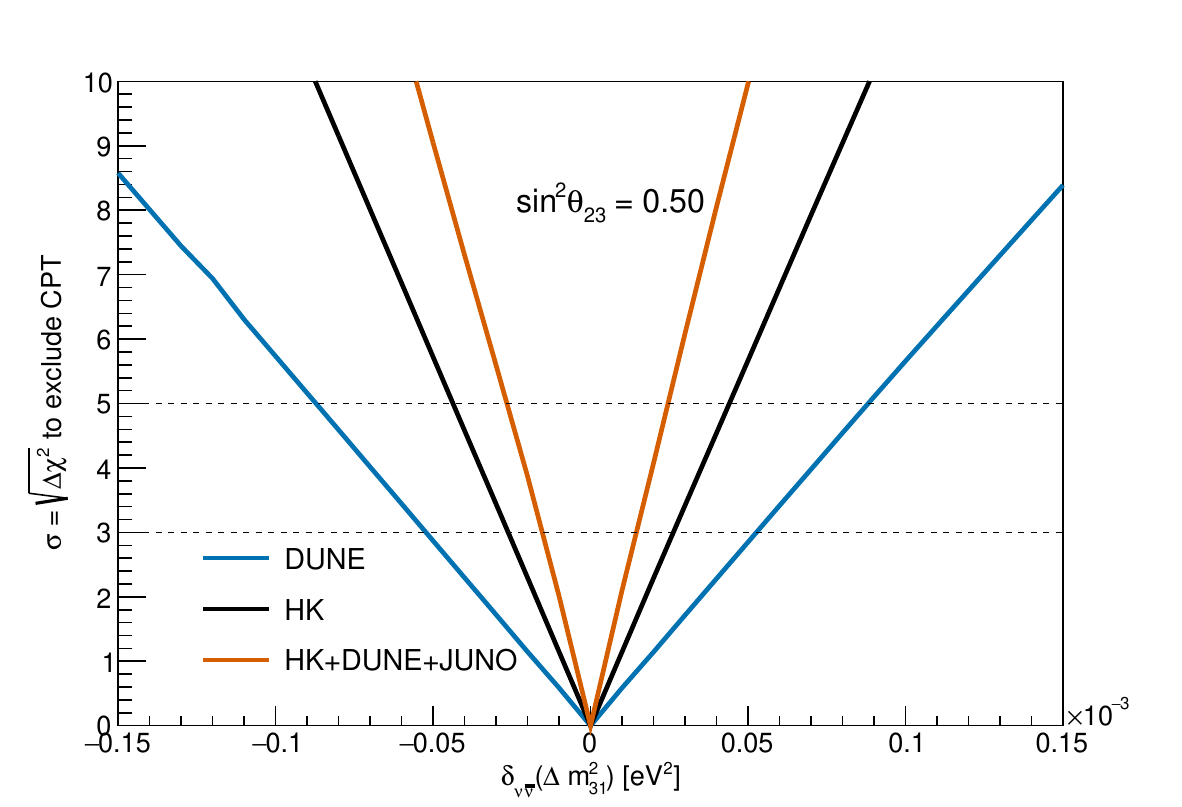}
\caption{\label{fig:cptsensicomparedunehkjuno} The sensitivities to $CPT$ violation as a function of \textit{true} \dcptdm for DUNE (blue), Hyper-K (black), and Hyper-K+DUNE+JUNO (vermilion) at $\sin^2\theta_{23} = \sin^2\overline{\theta}_{23} = 0.50$. $\Delta m^2_{31} 2.55\times 10^{-3}~\text{eV}^2$ is assumed to be \textit{true}.}
\end{figure}

\begin{table*}[htbp]
    \centering
    \begin{tabular}{c|c|c|c}
    \hline
    \hline
     $\sin^2\theta_{23} = \sin^2\overline{\theta}_{23}$ & DUNE  & HK & HK+DUNE+JUNO \\
     & [$eV^2$]& [$eV^2$] & [$eV^2$] \\
    \hline
      0.40 & $6.04\times 10^{-5}$ & $3.05\times 10^{-5}$ & $1.63\times 10^{-5}$ \\
      0.50 & $5.27\times 10^{-5}$ & $2.63\times 10^{-5}$ & $1.53\times 10^{-5}$ \\
      0.62 & $6.05\times 10^{-5}$ & $3.03\times 10^{-5}$ & $1.73\times 10^{-5}$ \\
      \hline
    \end{tabular}
    \caption{The lower limits for the \textit{true} $|\delta_{\nu\overline{\nu}}(\Delta m^2_{31})|$ to exclude $CPT$ conservation hypothesis at 3$\sigma$ CL. Assuming $\sin^2\theta_{23} = \sin^2\overline{\theta}_{23}$, values are calculated for three benchmark mixing scenarios (lower octant: 0.40, maximal mixing: 0.50, and higher octant: 0.62).}
    \label{tab:cptsensi}
\end{table*}

In this section, we investigate the potential of the combined Hyper-K, DUNE, and JUNO experiments to probe the $CPT$ violation in the leptonic sector. In particular, we aim to quantify the smallest difference between mass-squared splittings of neutrino and antineutrino that the experiments can detect the $CPT$ violation. We first examine the dependence of the expected sensitivity on the \textit{true} value of the atmospheric mixing angle $\theta_{23}$, and then compare the performance of the combined experimental configuration with that of the standalone Hyper-K and DUNE. The assumed neutrino-interaction models and their associated systematic uncertainties could affect the determination of the atmospheric mass-squared splittings, which in turn have an impact on the projected sensitivity. We therefore also investigate how a possible bias in the determination of the atmospheric mass-squared splitting could affect the sensitivity, and assess the level of bias control required to achieve sub-percent sensitivity. To quantify sensitivity, we fix the \textit{true} neutrino mass-squared splitting to $\Delta m^2_{31} = 2.55\times 10^{-3}~\text{eV}^2$ and scan the antineutrino parameter $\Delta \overline{m}^2_{31}$ in the interval from $\Delta m^2_{31} - 0.15\times 10^{-3}~\text{eV}^2$ to $\Delta m^2_{31} + 0.15\times 10^{-3}~\text{eV}^2$, with a uniform step of $0.01 \times 10^{-3}~\text{eV}^2$. At each point, we calculate the statistical sensitivity to the $CPT$-violating parameter \dcptdm, using the $\chi^2$ analysis described in the previous section. Repeating the scan for different true values of $\sin^2\theta_{23}$ (0.40, 0.50, and 0.62) allows us to determine how the sensitivity varies across the considered mixing-angle range, which is shown in Fig.~\ref{fig:cptsensidunehkjuno}. The results indicate that the sensitivity depends only weakly on the assumed mixing angle within the considered range of $0.40 \leq \sin^2\theta_{23} \leq 0.62$. The best sensitivity is obtained near maximal mixing, while a slightly weaker sensitivity is observed toward the lower and upper octants. 
Fig.~\ref{fig:cptsensicomparedunehkjuno} compares the sensitivities of DUNE, Hyper-K, and their combination with JUNO at $\sin^2\theta_{23}=0.50$. The numerical results are summarized in Table~\ref{tab:cptsensi}. At $3\sigma$ CL, the combined configuration of the three experiments can probe \dcptdm down to approximately $1.5 - 1.7 \times10^{-5}~\text{eV}^2$, depending on the \textit{true} value of $\theta_{23}$. This represents a substantial improvement over the sensitivities achievable with DUNE or Hyper-K individually, demonstrating the complementarity of the three experiments.

These sensitivity projections are obtained using the nominal systematic uncertainties implemented in our analysis. However, possible deficiencies in the interaction models, which may not be fully covered by the assigned systematic uncertainties, could introduce biases in the determination of oscillation parameters. In long-baseline accelerator experiments, uncertainties in neutrino–nucleus interactions can affect the reconstructed energy spectrum which directly affect the measurments of $\Delta m^2_{31}$ and $\Delta \overline{m}^2_{31}$. This issue has been investigated in the T2K simulated-data studies \cite{T2K20233d6e21pot}. The motivation of those studies is to examine whether the nominal interaction model and its associated systematic uncertainties can adequately accommodate possible variations in neutrino-interaction physics. By generating simulated data under alternative interaction assumptions and analyzing them with the nominal model, the studies assess the resulting shifts in the extracted oscillation parameters. Among the interaction-related effects considered in the T2K studies, pion secondary interactions make the largest individual contribution to the estimated systematic uncertainty of the atmospheric mass-squared splitting. The overall contributions from the simulated-data studies gives an effective smearing of $1.35\times10^{-5}~\text{eV}^2$ in the $\Delta m^2_{32}$. To investigate the implications of this effect for the present $CPT$-sensitivity analysis, we adopt this benchmark bias and assume that the bias magnitudes of $\Delta m^2_{31}$ and $\Delta m^2_{32}$ are approximately equal. In our calculation, the possible bias is incorporated by evaluating the $\chi^2$ for three values of the atmospheric mass-squared splitting: the nominal value and two shifted values corresponding to positive and negative bias. The minimum of these three $\chi^2$ values is then used in the subsequent sensitivity calculation. 

It is found that the impact of this benchmark bias is significant. For the combined configuration, the $3\sigma$ sensitivity threshold increases from $1.73\times10^{-5}~\text{eV}^2$ in the absence of bias to $3.06\times10^{-5}~\text{eV}^2$ when the benchmark bias is included. The later corresponds to $1.2\%$ relative difference compared to the assumed \textit{true} value $\Delta m^2_{31}=2.55\times10^{-3}~\text{eV}^2$. Thus, although the combined experiments offer excellent nominal sensitivity to $CPT$ violation, the impact of neutrino-interaction uncertainties could prevent the projected sensitivity from reaching the sub-percent precision. We therefore estimate the maximum tolerable bias (denoted by $b$) required to retain a sub-percent sensitivity. For the assumed \textit{true} value of $\Delta m^2_{31}$, this condition requires the $3\sigma$ sensitivity threshold to satisfy
$$ \delta_{\nu\bar\nu}(\Delta m^2_{31}) \leq 0.01\times\Delta m^2_{31} = 2.55\times 10^{-5}~\text{eV}^2.$$
A linear interpolation and simulated study between the no-bias case and benchmark-bias case give an estimate of $b \leq 0.83 \times 10^{-5}~\text{eV}^2$. This corresponds to a reduction of about $39\%$ relative to the benchmark value adopted from the T2K simulated-data studies. 
These results highlight the importance of controlling neutrino-interaction uncertainties in future precision oscillation experiments. In particular, reducing the impact of pion secondary interactions, together with improving the modeling and constraints on other relevant interaction processes, will be important for limiting possible biases in the extracted atmospheric mass-squared splitting. The ability to control such effects will be essential for fully exploiting the $CPT$ sensitivity of the combined analysis of Hyper-K, DUNE, and JUNO experiments.

We further examine how the $3\sigma$ sensitivity depends on the \textit{true} value of $\Delta m^2_{31}$, keeping the benmark bias fixed at $1.35\times 10^{-5}~\text{eV}^2$. We consider three values: the central value $\Delta m^2_{31} = 2.55 \times 10^{-3}~\text{eV}^2$, lower $1\sigma$ limit $2.49 \times 10^{-3}~\text{eV}^2$, and upper $1\sigma$ limit $2.60 \times 10^{-3}~\text{eV}^2$. The corresponding $3\sigma$ sensitivity thresholds are $3.06 \times 10^{-5}~\text{eV}^2$, $3.09 \times 10^{-5}~\text{eV}^2$, and $3.10 \times 10^{-5}~\text{eV}^2$. Relative to the central-value result, the $\pm 1\sigma$ shifts differ by approximately $1.0\%$ and $1.3\%$, respectively. This indicates that the main conclusions are not strongly dependent on the particular \textit{true} value adopted for $\Delta m^2_{31}$ over the considered $1\sigma$ range.

Overall, our studies demonstrate that the Hyper-K, DUNE, and JUNO combination has the potential to probe $CPT$ violation in the lepton sector at the sub-percent level, provided that the effective bias associated with neutrino-interaction uncertainties can be sufficiently controlled.

%%%%%%%%%%%%%%%%%%%%%%%%%%%%%%%%%%%%%%%%%%%%%%%%%%%%%%%%%%%%%%%%%%%
%%%%%%%%%%%%%%%%%%%%%%%%%%%%%%%%%%%%%%%%%%%%%%%%%%%%%%%%%%%%%%%%%%%
\section{\label{sec:fin}Conclusion}
%%%%%%%%%%%%%%%%%%%%%%%%%%%%%%%%%%%%%%%%%%%%%%%%%%%%%%%%%
%%%%%%%%%%%%%%%%%%%%%%%%%%%%%%%%%%%%%%%%%%%%%%%%%%%%%%%%%
In this study, we have conducted a sensitivity study of current and next-generation long-baseline neutrino experiments to test $CPT$ symmetry by examining the atmospheric mass-squared splittings $\Delta m^2_{31}$ and $\Delta \overline{m}^2_{31}$. By introducing a parameter $\delta_{\nu\overline{\nu}}(\Delta m^2_{31})$, we simulated and compared the expected sensitivity of DUNE, Hyper-K, and JUNO under various experimental configurations. Our study shows that future experiments like Hyper-K and DUNE, especially when combined with JUNO, can significantly improve the $CPT$ sensitivity, reaching as low as $\sim 1.7 \times 10^{-5}~\text{eV}^2$ at 3$\sigma$ CL under the assumption of $CPT$ invariance and fixed $\sin^2\theta_{23}=\sin^2\overline{\theta}_{23}$, which is at the same level as the solar sector. This represents an improvement of approximate $70\%$ over the combined sensitivity of T2K, NO$\nu$A, and JUNO. We have also investigated the dependence of the sensitivity on the atmospheric mixing angle $\theta_{23}$ and found that the sensitivity is optimized near maximal mixing, while moderately decreased toward either octant. Nevertheless, the configurations maintain a relatively stable sensitivity across the experimentally allowed $3\sigma$ range of $\sin^2\theta_{23}$.

An important aspect of this study is assessing potential biases in the determination of the atmospheric mass-squared splitting from uncertainties in neutrino-nucleus interaction models. Using the effective bias motivated by the T2K simulated-data studies, we find that the $3\sigma$ sensitivity threshold of the combined Hyper-K, DUNE, and JUNO increases from $1.73\times10^{-5}~\text{eV}^2$ to $3.06\times10^{-5}~\text{eV}^2$ - a relative difference of $1.2\%$ from the assumed \textit{true} atmospheric mass-squared splitting. This indicates that interaction-induced biases could significantly compromise the projected sub-percent sensitivity. Further estimation shows that in order to retain a sub-percent sensitivity to $CPT$ violation, the effective bias must be reduced to below $0.83\times10^{-5}~\text{eV}^2$, corresponding to an improvement of approximately $39\%$ relative to the T2K simulated benchmark. These findings emphasize the importance of improving neutrino interaction models and constraining the associated systematic effects in future precision oscillation measurements. We have also investigated the robustness of our conclusions against variations in the assumed \textit{true} value of $\Delta m^2_{31}$. We find that the resulting $3\sigma$ sensitivity thresholds vary by only around $1\%$ relative to the central-value result, demonstrating that our conclusions are robust within the $\pm 1\sigma$ range of \textit{true} $\Delta m^2_{31}$.

\vspace{1cm}
%%%%%%%%%%%%%%%%%%%%%%%%%%%%%%%%%%%%%%%%%%%%%%%%%%%%%%%%%
%%%%%%%%%%%%%%%%%%%%%%%%%%%%%%%%%%%%%%%%%%%%%%%%%%%%%%%%%
\section*{Acknowledgments}
%%%%%%%%%%%%%%%%%%%%%%%%%%%%%%%%%%%%%%%%%%%%%%%%%%%%%%%%%
%%%%%%%%%%%%%%%%%%%%%%%%%%%%%%%%%%%%%%%%%%%%%%%%%%%%%%%%%
This work is funded by the National Foundation for
Science and Technology Development (NAFOSTED) of
Vietnam under Grant No. 103.99-2023.144.

%%%%%%%%%%%%%%%%%%%%%%%%%%%%%%%%%%%%%%%%%%%%%%%%%%%%%%%%%
%%%%%%%%%%%%%%%%%%%%%%%%%%%%%%%%%%%%%%%%%%%%%%%%%%%%%%%%%
%\bibliographystyle{apsrev} 
%\bibliographystyle{ieeetr} 
\bibliographystyle{jhep} 
\bibliography{apssamp}% Produces the bibliography via BibTeX.
%%%%%%%%%%%%%%%%%%%%%%%%%%%%%%%%%%%%%%%%%%%%%%%%%%%%%%%%%%%%%%%%%%%%%%%%%%%%%%%%%%%%%%%%%%%%%%%%%%%%%%%%%%%%%%%%%%
\end{document}